\documentclass[11pt]{cernrep}

\usepackage{graphicx}
\usepackage{here}

\newcommand{\lsim}{\,{\buildrel < \over {_\sim}}\,}
\newcommand{\gsim}{\,{\buildrel > \over {_\sim}}\,}

\begin{document}

\title{
Nonlinear corrections to the DGLAP equations; looking for the
saturation limits\footnote{~~Contribution to CERN Yellow Report on
Hard Probes in Heavy Ion Collisions at the LHC.}}

\author{K.J. Eskola $^{\rm a,b,}$,
 H. Honkanen $^{\rm a,b}$,
 V.J. Kolhinen $^{\rm a,b}$,
 Jianwei Qiu $^{\rm c}$,
 C.A. Salgado $^{\rm d}$}

\institute{
{\em $^{\rm a}$ Department of Physics, 
P.O.Box 35, FIN-40014 University of Jyv\"askyl\"a, Finland\\}
{\em $^{\rm b}$ Helsinki Institute of Physics,
P.O.Box 64, FIN-00014 University of Helsinki, Finland\\}
{\em $^{\rm c}$ Department of Physics and Astronomy, 
Iowa State University, Ames, Iowa, 50011, U.S.A.\\}
{\em $^{\rm d}$ CERN, Theory Division, CH-1211 Geneva, Switzerland}
}

\maketitle

\vspace{-5.4cm}
\begin{flushright}
  HIP-2003-08/TH
\end{flushright}
\vspace{4.5cm}

\begin{abstract}
The effects of the first nonlinear corrections to the DGLAP equations
are studied in light of the HERA data. Saturation limits are
determined in the DGLAP+GLRMQ approach for the free proton and for the
Pb nucleus.
\end{abstract}

%
%

\section{Introduction}

Parton distribution functions (PDF) of the free proton, $f_i(x,Q^2)$,
are needed for the calculation of the cross sections of hard processes
in hadronic collisions. Once they are determined at certain
initial scale $Q_0^2$, the DGLAP equations \cite{Gribov:ri} describe
well their scale evolution at large scales. Based on the global fits
to the available data several different parametrizations of PDF have
been obtained \cite{Martin:2001es,Lai:1999wy,Pumplin:2002vw}.  The
older PDF sets do not describe adequately the recent HERA data
\cite{Adloff:2000qk} on the structure function $F_2$ at the
perturbative scales $Q^2$ at small $x$.  In the analysis of newer PDF
sets, such as CTEQ6 \cite{Pumplin:2002vw} and MRST2001
\cite{Martin:2001es}, these data have been taken into
account. However, difficulties arise when fitting both small and large
scale data simultaneously. In the MRST set, the entire H1 data set
\cite{Adloff:2000qk} has been used in the analysis, leading to a good
average fit at all scales, but at the expense of allowing for a
negative NLO gluon distribution at small $x$ and $Q^2\lsim 1$ GeV$^2$.
In the CTEQ6 set only the large scale ($Q^2>4$ GeV$^2$) data have been
included, giving a good fit at large $Q^2$, but leaving the fit at
small-$x$ and small $Q^2$ ($Q^2<4$ GeV$^2$) region worse. Moreover,
the gluon distribution at the values of small $x$ and $Q^2 \lsim 1.69$
GeV$^2$ has been set to zero.

These problems are interesting as they can be signs of a new QCD
phenomenon: at small values of momentum fraction $x$ and scales $Q^2$,
gluon recombination terms, which lead to nonlinear corrections to the
evolution equations, can become significant.  First of these nonlinear
terms have been calculated by Gribov, Levin and Ryskin
\cite{Gribov:tu}, and, Mueller and Qiu \cite{Mueller:wy}.  In the
following these correction terms shall be referred to as GLRMQ terms for
short. With the modifications, the evolution equations become \cite{Mueller:wy}
\begin{eqnarray}
\frac{\partial xg(x,Q^2) }{\partial \log Q^2} 
  &=&   \frac{\partial xg(x,Q^2) }{\partial \log Q^2}\bigg|_{\rm DGLAP} 
   - \quad \frac{9\pi}{2} \frac{\alpha_s^2}{Q^2} 
    \int_x^1 \frac{dy}{y} y^2 G^{(2)}(y,Q^2), \label{gl-evol} \\
\frac{\partial x\bar{q}(x,Q^2)}{\partial \log Q^2}   & = & 
\frac{\partial x\bar{q}(x,Q^2)}{\partial \log Q^2}\bigg|_{\rm DGLAP}
  -  \quad \frac{3\pi}{20}\frac{\alpha_s^2}{Q^2} 
     x^2 G^{(2)}(x,Q^2)
   + \ldots G_{\rm HT}, \label{sea-evol}
\end{eqnarray} 
where the two-gluon density can be modelled as $ x^2G^{(2)}(x,Q^2)=
\frac{1}{\pi R^2}[xg(x,Q^2)]^2, $ with the radius of the proton
$R=1$~fm.  The higher dimensional gluon term $ G_{\rm HT}$
\cite{Mueller:wy} is here assumed to be zero.  The effects of the
nonlinear corrections to the DGLAP evolution of the PDF of the free
proton were studied in \cite{Eskola:2002yc} in view of the recent H1
data; the results are discussed below.

\section{The analysis}

\begin{figure}[tbh]
\begin{center}
\includegraphics[width=8cm]{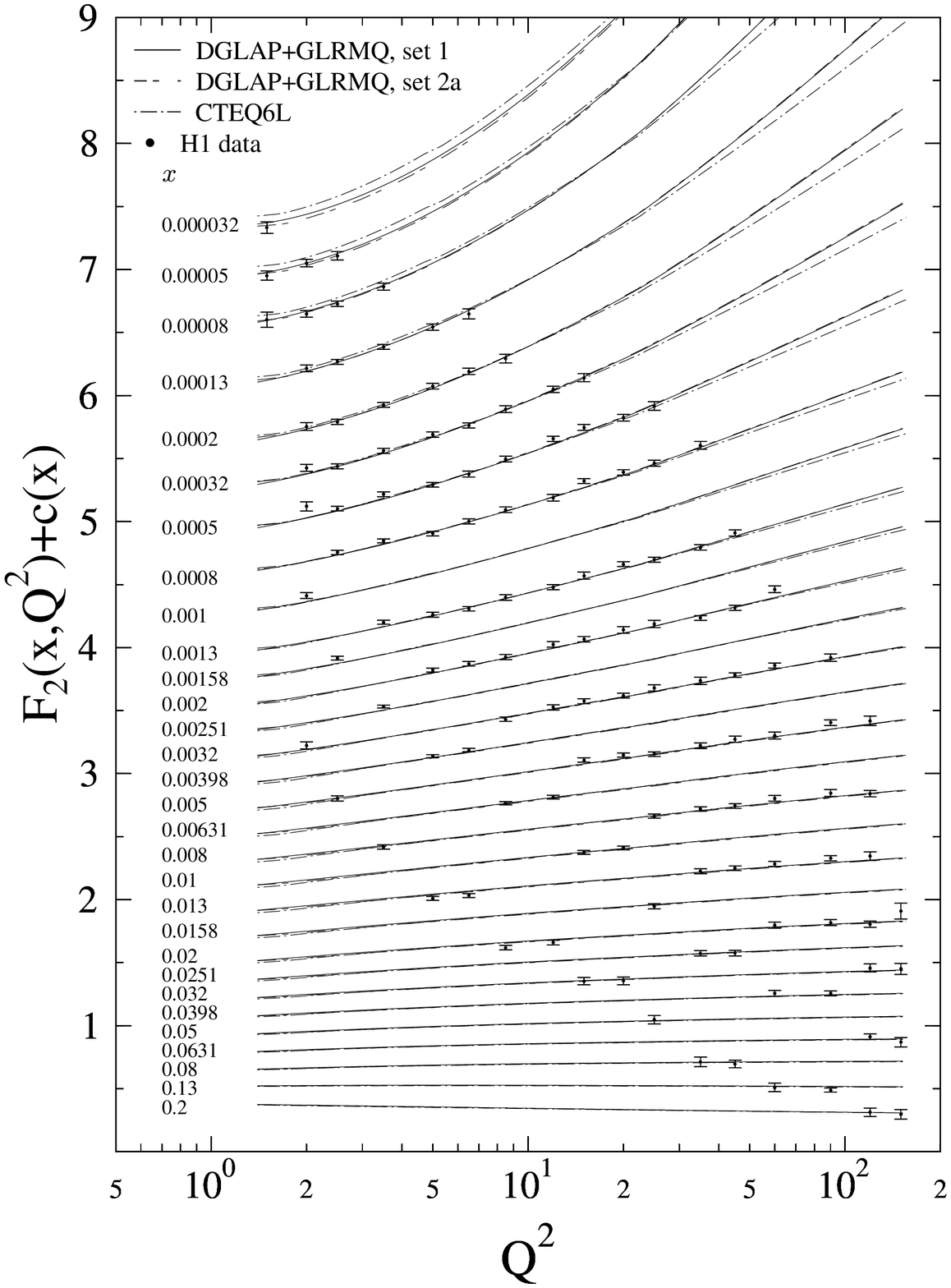}
\includegraphics[width=6.9cm]{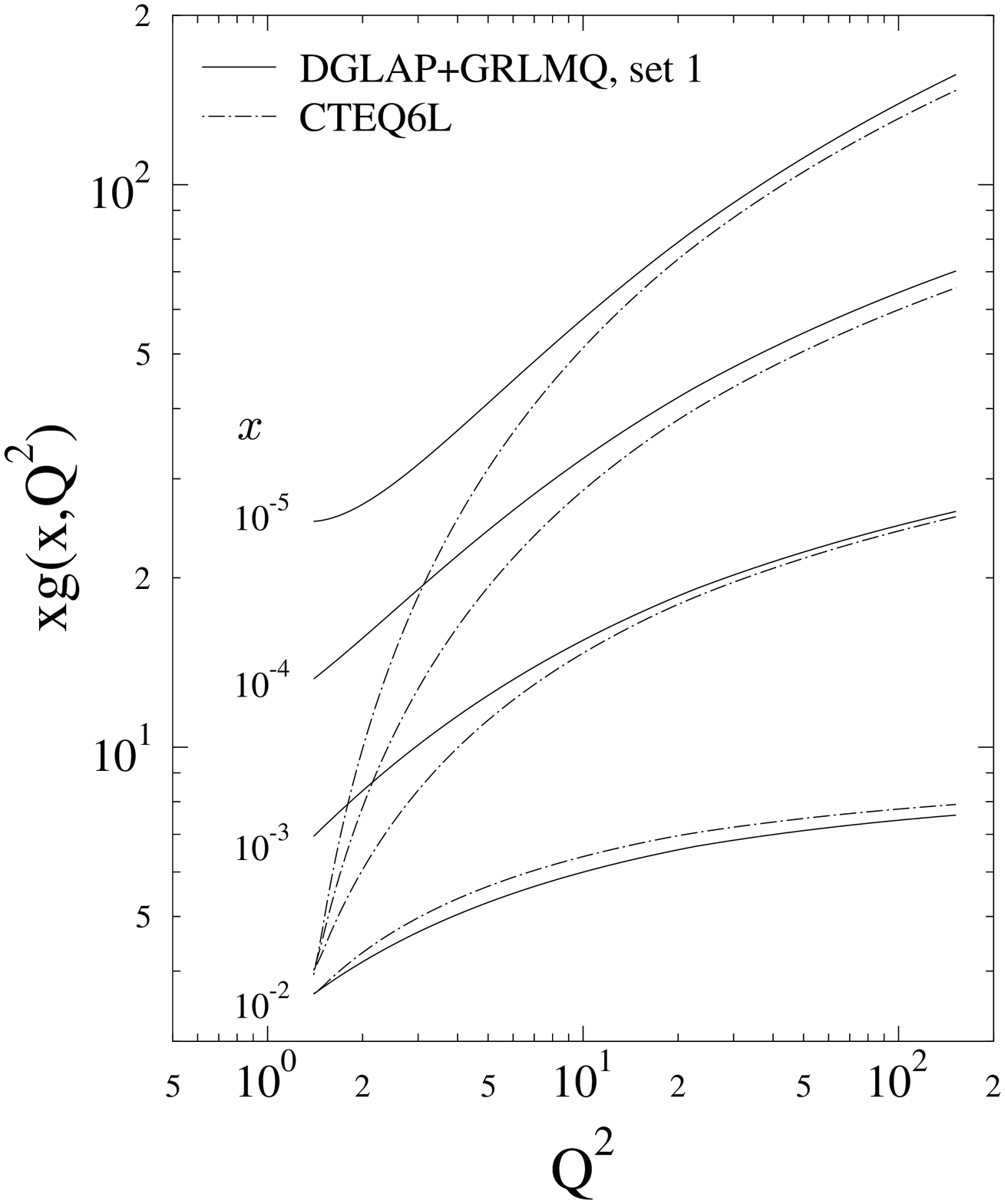} 
\caption[a]{{\small {\bf Left:} $F_2(x,Q^2)$ calculated using
CTEQ6L \cite{Pumplin:2002vw} (dotted-dashed) and the DGLAP+GLRMQ
results with set 1 (solid) and set 2a (double dashed)
\cite{Eskola:2002yc}, compared with the H1 data
\cite{Adloff:2000qk}. {\bf Right:} The $Q^2$ dependence of the gluon
distribution function at fixed $x$, from set 1 evolved with
DGLAP+GLRMQ (solid), and directly from CTEQ6L (dotted-dashed).} }
\label{F2_vs_cteq6}
\end{center}
\end{figure}

The goal of the analysis in \cite{Eskola:2002yc} was (1) to possibly
improve the (LO) fit of the calculated $F_2(x,Q^2)$ to the H1 data
\cite{Adloff:2000qk} at small $Q^2$, while (2) at the same time
maintain the good fit at large $Q^2$, and finally (3) to study the
interdependence between the initial distributions and the evolution.

In CTEQ6L a good fit to the H1 data is obtained (see
Fig.~\ref{F2_vs_cteq6}) with a flat small-$x$ gluon distribution at
$Q^2\sim 1.4$~GeV$^2$.  As can be seen from
Eqs.~(\ref{gl-evol}-\ref{sea-evol}), the GLRMQ corrections slow down
the scale evolution. Now one may ask whether the H1 data can be
reproduced equally well with different initial conditions
(i.e. assuming larger initial gluon distributions) and the GLRMQ
corrections included in the evolution.  This question has been studied
in \cite{Eskola:2002yc} by generating three new sets of initial
distributions using DGLAP + GLRMQ evolved CTEQ5 \cite{Lai:1999wy} and
CTEQ6 distributions as guidelines.  The initial scale was chosen to be
$Q_0^2=1.4$ GeV$^2$, slightly below the smallest scale of the data
points.  The modified distributions at $Q_0^2$ were constructed
piecewise from CTEQ5L and CTEQ6L distributions evolved down from $Q^2$
= 3 and 10 GeV$^2$ (CTEQ5L) and $Q^2$ = 5 GeV$^2$ (CTEQ6L). A power
law form was used in the small-$x$ region to tune the initial
distributions until a good agreement with the H1 data was found.

The difference between the three sets in \cite{Eskola:2002yc} is that in
set 1 there is still a nonzero charm distribution at 
$Q_0^2=1.4$ GeV$^2$, which is slightly below the charm mass treshold,
taken to be $m_c=1.3$ GeV in CTEQ6.  In sets 2 the charm distribution has
been removed at the initial scale and the resulting deficit in $F_2$
has been compensated by slightly increasing the other sea quarks at
small $x$. Moreover, the effect of the charm was studied by using
different mass tresholds: $m_{\rm c}=1.3$ GeV in set 2a whereas in set
2b it is $m_{\rm c}=\sqrt{1.4}$ GeV, i.e. charm begins to evolve
immediately from the initial scale.

The results from the DGLAP+GLRMQ evolution with the new initial
distributions are shown in Figs.~\ref{F2_vs_cteq6}. The left panel
shows the comparison between the H1 data and the (LO) structure
function $F_2(x,Q^2)=\sum_i e_i^2 x[q_i(x,Q^2)+\bar q_i(x,Q^2)]$
calculated from set 1 (solid lines), set 2a (double dashed) and the
CTEQ6L parametrization (dotted-dashed lines). As can be seen, the
results are very similar, which shows that with modified initial
conditions and DGLAP+GLRMQ evolution, one obtains as good or even a
better fit to the HERA data ($\chi/N = 1.13$, 1.17, 0.88 for the sets
1, 2a, 2b, correspondingly) as with the CTEQ6L distributions ($\chi/N
= 1.32$).

The evolution of the gluon distribution functions in the DGLAP+GLRMQ
and DGLAP cases is illustrated more explicitly in the right panel of
Fig. \ref{F2_vs_cteq6}, in which the absolute distributions for fixed
$x$ are plotted as a function of $Q^2$ for set~1 and for CTEQ6L. The
figure shows how the differences which are large at initial scale
vanish during the evolution due to the GLRMQ effects.  At scales $Q^2
\gsim 4$~GeV$^2$ the GLRMQ corrections fade out rapidly and the DGLAP
terms dominate the evolution.

\section{Saturation}

The DGLAP+GLRMQ approach also offers a way to study the gluon
saturation limits. For each $x$ in the small-$x$ region, the
saturation scale $Q_{\rm sat}^2$ can be defined as the value of the
scale $Q^2$ where the DGLAP and GLRMQ terms in the nonlinear evolution
equation become equal, $\frac{\partial xg(x,Q^2)}{\partial \log
Q^2}|_{Q^2=Q_{\rm sat}^2(x)}=0$.  The region of applicability of the
DGLAP+GLRMQ is at $Q^2>Q^2_{\rm sat}(x)$ where the linear DGLAP part
dominates the evolution.  In the saturation region, at $Q^2<Q^2_{\rm
sat}(x)$, the GLRMQ terms dominate, and all nonlinear terms become
important.

\begin{figure}[htb]
\begin{center}
\includegraphics[width=7.5cm]{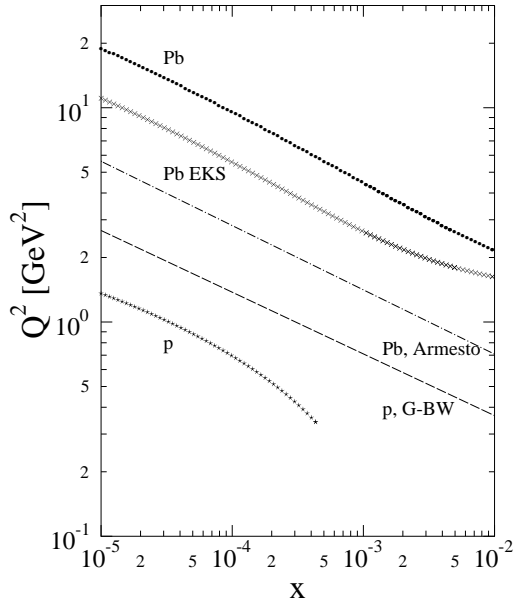} 
\caption{{\small The gluon saturation limits in the DGLAP+GLRMQ approach 
for proton (asterisks) and Pb ($A=208$), with (crosses) and 
without (dots) nuclear modifications \cite{Eskola:2002yc}.  The
saturation line for the proton from the geometric saturation model 
\cite{Golec-Biernat:1998js} (dashed line), and for Pb from \cite{Armesto:2002ny} (dotted-dashed) are also plotted.
 }}
\label{satur}
\end{center}
\end{figure}

In order to find the saturation scales $Q^2_{\rm sat}(x)$ for the free
proton, the obtained initial distributions (set 1) at
$Q_0^2=1.4$~GeV$^2$ have to be evolved downwards in scale using the
DGLAP+GLRMQ equations.  As discussed in \cite{Eskola:2002yc}, since
only the first correction term has been taken into account, the gluon
distribution near the saturation region should be considered as an
upper limit. Consequently, the obtained saturation scale is an upper
limit as well. The result is shown in Fig. \ref{satur} (asterisks).
The saturation line for the free proton from the geometric saturation
model by Golec-Biernat and W\"usthoff (G-BW)
\cite{Golec-Biernat:1998js} is also plotted (dashed line) for
comparison. It is interesting to note that although the DGLAP+GLRMQ
and G-BW approaches are very different, the slopes of the curves are
very similar at the smallest values of $x$.

Saturation scales for nuclei can also be determined in a similar
manner.  For a nucleus $A$, the two-gluon density can be modelled as
$x^2G^{(2)}(x,Q^2)= \frac{A}{\pi R_A^2}[xg(x,Q^2)]^2$, i.e. the effect
of the correction is enhanced by a factor of $A^{1/3}$. Now a first
estimate for the saturation limit can be obtained by starting the
downwards evolution at high enough scales, $Q^2=100 \ldots 200$
GeV$^2$, where the GLRMQ terms are negligible. The result, which
similarly to the proton case is an upper limit, is shown for Pb in
Fig.~\ref{satur} (dots).  The effect of the nuclear modifications was
also studied by applying the EKS98 \cite{Eskola:1998df}
parametrization at the high starting scale.  As a result, the
saturation scales $Q_{\rm sat}^2(x)$ are somewhat reduced, as shown in
Fig.~\ref{satur} (crosses).  The saturation limit obtained for a Pb
nucleus by Armesto in a Glauberized geometric saturation model
\cite{Armesto:2002ny} is shown (dotted-dashed) for comparison. Again,
despite of the differences between the approaches, the slopes of the
curves are strikingly similar.

For further studies and for more accurate estimates of $Q_{\rm
sat}^2(x)$ in the DGLAP+GLRMQ approach, a full global fit analysis for
the nuclear parton distribution functions should be performed, along
the same lines as in EKRS \cite{Eskola:1998iy,Eskola:1998df} and in
HKM \cite{Hirai:2001np}.

\vspace{1cm}
\noindent{\large \bf Acknowledgements.}  We thank N. Armesto,
P.V. Ruuskanen, I. Vitev and other participants of the CERN Hard
Probes workshop for discussions. We are grateful for the Academy of
Finland, Project 50338, for financial support.  J.W.Q. is supported in
part by the United States Department of Energy under Grant
No. DE-FG02-87ER40371.  C.A.S. is supported by a Marie Curie
Fellowship of the European Community programme TMR (Training and
Mobility of Researchers), under the contract number
HPMF-CT-2000-01025.

\end{document}